\documentclass[%
preprint,
pra,
amsmath,amssymb,
aps,
]{revtex4-1}
\usepackage{bm}
\usepackage{graphicx}
\usepackage[caption=false]{subfig}
\usepackage{braket}
\newcommand{\defas}{\equiv}
\graphicspath{{./fig/}}
\usepackage{color}

\begin{document}

\title{Excited state search using quantum annealing}
\author{Yuya Seki}
\author{Yuichiro Matsuzaki}
\author{Shiro Kawabata}
\affiliation{Nanoelectronics Research Institute, National Institute of Advanced Industrial Science and Technology (AIST), 1-1-1 Umezono, Tsukuba, Ibaraki 305-8568 Japan}
\date{\today}

\begin{abstract}
Quantum annealing (QA) is one of the ways to search the ground state of the problem Hamiltonian.
Here, we propose the QA scheme to search \textit{arbitrary} excited states of the problem Hamiltonian.
In our scheme, an $n$-th excited state of the trivial Hamiltonian is initially prepared and is adiabatically changed into an $n$-th excited state of the target Hamiltonian.
Although our scheme is general such that we can search any excited states,
we especially discuss the first excited state search in this paper.
As a comparison, we consider a non-adiabatic scheme to find the first excited state
with non-adiabatic transitions from the ground state.
By solving the Lindblad master equation, we evaluate the performance of each scheme
under the influence of decoherence.
Our conclusion is that the adiabatic scheme show better performance than the non-adiabatic scheme
as long as the coherence time of qubits is sufficiently long.
These results are important for applications in the area of quantum chemistry,
quantum simulation, and post-quantum cryptography.
\end{abstract}

\maketitle

\section{\label{sec:intr}Introduction}

Exploring new applications of quantum devices is an important challenge to be addressed
for the development of quantum technologies.
In recent years, a lot of groups have put intensive efforts into developing devices
for computing using quantum effects.
For example, D-Wave Systems, Inc.\ has developed quantum devices~\cite{johnson2011quantum} based on quantum annealing (QA) and adiabatic quantum computation (AQC)~\cite{kadowaki1998quantumannealing,farhi2000quantum,farhi2001aquantum},
and released D-Wave 2000Q consisting of over two thousands of superconducting qubits~\cite{dwave2000q}.
In addition, several quantum devices for QA have been proposed and developed~\cite{barends2016digitized,rosenberg20173dintegrated,maezawa2019toward,novikov2018exploring,mukai2019superconducting}.
Although it is unclear if such devices can resolve real world important issues,
we cloud still use such devices to check whether these really provide us with the solutions
of practical problems much faster than existing classical computers.
For this purpose, it is important to look for the applications of QA.

Quantum chemistry calculations can be one of possible important applications for the QA devices.
The statical and dynamical properties of molecules studied in quantum chemistry are obtained
from the second-quantized Hamiltonians for molecules.
It requires exponentially large amount of memory to apply ab initio calculations on classical computers to the quantum chemistry calculations, since the dimension of the Fock space grows exponentially as the number of the modes increases.
Importantly, QA can be used to analyze the second-quantized Hamiltonians by using results from the following studies.
Bravyi and Kitaev have developed a method to transform the second-quantized many-body Hamiltonians
to those of spin-1/2 systems, where the number of the spins is the same as the that of modes~\cite{bravyi2002fermionic,seeley2012bravyi,tranter2015bravyi}.
A more practical representation of the molecular systems that includes only 2-local interactions
between the spins has been developed in Ref.~\cite{babbush2014adiabatic}.
Furthermore, Xia \textit{et al.}\ have found a approximated representation of the molecular systems
that can be analyzed using the current D-Wave machines~\cite{xia2017electronic}.
Streif \textit{et al.}\ have applied the D-Wave machine to calculate the ground state energy
of molecular systems on the basis of the above studies,
and succeeded in obtaining an estimation of the ground state energy
that is close to the exact solution of molecular hydrogen $\mathrm{H}_2$
as well as Lithium hydride $\mathrm{LiH}$ on the D-Wave 2000Q~\cite{streif2019solving}.

Excited state search is of great importance as well as the ground state search.
For example, excited states are related to important physical properties of molecular systems~\cite{serrano-andres2005quantumchemistry}.
In order to explain (or predict) the spectroscopic and photochemical properties of molecules,
the information on the excited states as well as the ground state is required.
Excited state search also appears in post-quantum cryptography~\cite{bernstein2017post}.
The post-quantum cryptography is an important technology to ensure secure communication
that is robust to fault tolerant quantum computers.
Some of the post-quantum cryptography base on the shortest vector problem, which belongs to the class of NP-hard problems~\cite{ajtai1998shortest}.
Joseph \textit{et al.}\ have investigated the performance of QA for the shortest vector problem,
where the solution of the problem is embedded in the first excited state of the problem Hamiltonian~\cite{joseph2019notsoadiabatic}.
In their method, a ground state of an initial Hamiltonian is prepared, then let the state evolve according to the Shr\"{o}dinger equation.
They numerically obtained the desired first excited state through non-adiabatic transitions
from the ground to the first excited state.

\begin{figure}[htbp]
    \centering
    \includegraphics[width=3.2truein]{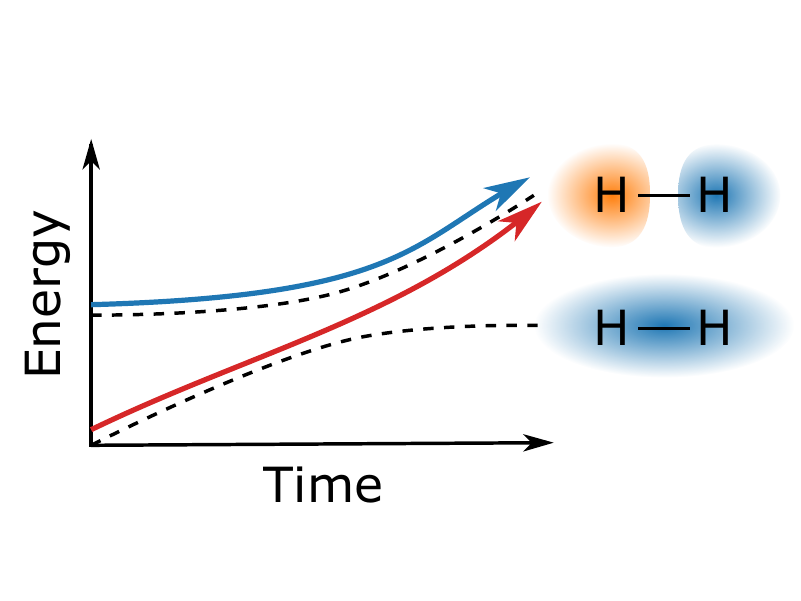}
    \caption{Schematic view of the adiabatic and non-adiabatic schemes to find the first excited state.
    The dashed lines represent instantaneous energy eigenvalues of a system at each time.
    The blue arrow following the first excited state describes the evolution of quantum state
    in the adiabatic scheme,
    and the red arrow from the ground state to the first excited state
    corresponds to the non-adiabatic scheme.
    The right figures show the ground state of the molecular hydrogen $\mathrm{H}_2$, the bonding orbital (bottom), and the first excited state of the antibonding orbital (top).
    }
    \label{fig:schemes}
\end{figure}

In this paper, we propose an adiabatic scheme to search $n$-th excited states
by the adiabatic evolution.
In addition, we also compare two schemes, namely adiabatic and non-adiabatic schemes,
for obtaining first excited states as shown in Fig.~\ref{fig:schemes}.
The main proposition of the present paper is that the adiabatic scheme proposed in this paper
has higher performance than the non-adiabatic scheme~\cite{joseph2019notsoadiabatic} as long as noise is weak.
In order to compare these two schemes, we apply them to the problem of finding
the first excited state of the spin-star model, which is analytically tractable
and has the similar type of interaction Hamiltonian for the molecular systems~\cite{hutton2004mediated}.
To take noise into account, we adopt the Lindblad master equation,
and compare these two schemes by numerically solving the master equation.
We find that the adiabatic scheme has a better performance than the non-adiabatic scheme
as long as the coherence time of the qubit is longer than the necessary evolution time
to satisfy the adiabatic condition.

Although the present paper investigates schemes based on QA,
it should be noted that quantum algorithms on the fault tolerant gate-type quantum computers
as well as noisy intermediate scale quantum computers
for quantum chemistry calculations has been actively developed~\cite{mcardle2018quantum}.
Quantum algorithms to calculate the properties of excited states has also been developed~\cite{nakanishi2019subspace,jones2019variational,higgott2019variational,jouzdani2019method}.
However, methods on QA to search excited states have not been studied well.

The remainder of this paper is organized as follows.
In Sec.~\ref{sec:method}, we introduce the adiabatic and non-adiabatic schemes
to obtain excited estates based on QA.
We show the Hamiltonian of the spin-star model in Sec.~\ref{sec:model}.
Section~\ref{sec:setup and methods} is devoted to explain protocol to apply the two schemes
to the spin-star model, and describing how to compare the schemes.
We show numerical results, and give an interpretation to the results in Sec.~\ref{sec:results}.
Finally, we conclude in Sec.~\ref{sec:conclusion}.

\section{\label{sec:method}Adiabatic and non-adiabatic schemes}

We discuss two schemes to obtain excited states of spin-1/2 systems, namely adiabatic scheme and non-adiabatic schemes.
The adiabatic scheme, which is known as adiabatic quantum computation,
is based on the adiabatic theorem~\cite{van2001powerful}.
Let us consider a problem of finding an $n$-th excited state of a problem Hamiltonian.
Assuming that there are no crossings for the $n$-th energy level of the total Hamiltonian
during the annealing process, we can obtain the $n$-th excited state
of the problem Hamiltonian with a high fidelity by evolving the $n$-th excited state
of the driver Hamiltonian sufficiently slowly in the absence of noise.
Under noisy environment, as we increase the evolution time,
the decoherence is more significant, resulting in a low fidelity.
Hence, there should be an optimal evolution time in order to obtain the highest fidelity
in the noisy environment.

The non-adiabatic scheme could provide us with a way to obtain the desired excited state
by inducing the non-adiabatic transitions from the ground state to the other ones.
In the present paper, we prepare the ground state of the trivial initial Hamiltonian
such as the conventional transverse-field term,
then non-adiabatically change the parameters of the Hamiltonian so that the final state
should have some overlap with the desired state due to the Landau--Zener transition.
It is worth mentioning that, the adiabatic scheme could provide arbitrary excited states
with a high fidelity without any decoherence due to the adiabatic theorem.
On the other hand, there is no guarantee for the non-adiabatic scheme to obtain the desired state
with a high fidelity.

\section{\label{sec:model}Spin-star model}
In this section, we introduce the spin-star model~\cite{hutton2004mediated}.
It is worth mentioning that this model contains both $XX$ and $YY$ interactions
unlike the conventional Ising type model that only contains $ZZ$ interactions.
Importantly, considerable attention has been attracted to such $XX$ and $YY$ interactions
in the field of QA, because the typical Hamiltonian in the quantum chemistry
(that is one of the practical applications in the QA) contains such terms.

\begin{figure}[htbp]
    \centering
    \includegraphics[width=3.2truein]{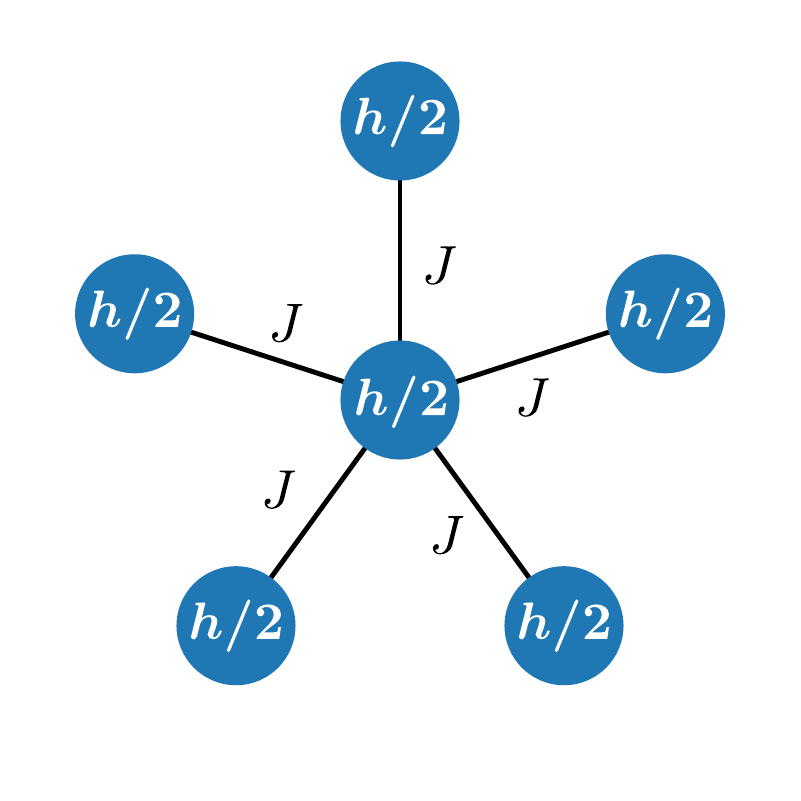}
    \caption{Schematic figure of the spin-star model with one central spin represented by the central circle and five satellite spins represented by the surrounding circles.
    The uniform longitudinal field, $h/2$, is subjected to each spin.
    The central spin is coupled with every satellite spin through spin flip-flop interactions  with strength $J$ depicted by the solid lines between the circles.
    }
    \label{fig:spin_star_model}
\end{figure}

We show the Hamiltonian of the spin-star model with a uniform longitudinal field,
i.e. without detuning.
The schematic figure of the spin-star model is shown in Fig.~\ref{fig:spin_star_model}.
Under the uniform longitudinal filed $h / 2$,
the Hamiltonian of the spin-star model is given by
\begin{align}
    \hat{H}_{\text{P}} = \frac{h}{2} \sum_{i=0}^{N} \hat{\sigma}_{i}^{z}
        + J\sum_{i=1}^{N}(\hat{\sigma}_{0}^{+}\hat{\sigma}_{i}^{-} + \hat{\sigma}_{0}^{-}\hat{\sigma}_{i}^{+}),
    \label{eq:target H}
\end{align}
where the spin at the zeroth site corresponds to the central spin
and the others to the satellite spins.
Throughout this paper, we set $\hbar = 1$.
The Pauli operators acting on site $i$ are denoted as $\sigma_{i}^{x}, \sigma_{i}^{y}$, and $\sigma_{i}^{z}$.
The creation and annihilation operators are $\sigma_{i}^{\pm} = \sigma_{i}^{x} \pm i \sigma_{i}^{y}$.
The Hamiltonian~\eqref{eq:target H} can be rewritten using the angular momentum coupling
of the satellite spins as
\begin{align}
    \hat{H}_{\text{P}} = \frac{h}{2} \hat{\sigma}_{0}^{z} + h \hat{S}^{z}
                 + J (\hat{\sigma}_{0}^{+}\hat{S}^{-} + \hat{\sigma}_{0}^{-}\hat{S}^{+}),
    \label{eq:target H with total spin}
\end{align}
where the $z$ component of the total angular momentum is defined
as $\hat{S}^{z} \defas \sum_{i=1}^{N}\hat{\sigma}_{i}^{z}/2$,
and the $x$ and $y$ components are defined similarly.
The spin ladder operators for the total angular momentum are defined as
$\hat{S}^{\pm} \defas \hat{S}^{x} \pm i \hat{S}^{y}$.

\begin{figure}[htbp]
    \centering
    \includegraphics[width=3.2truein]{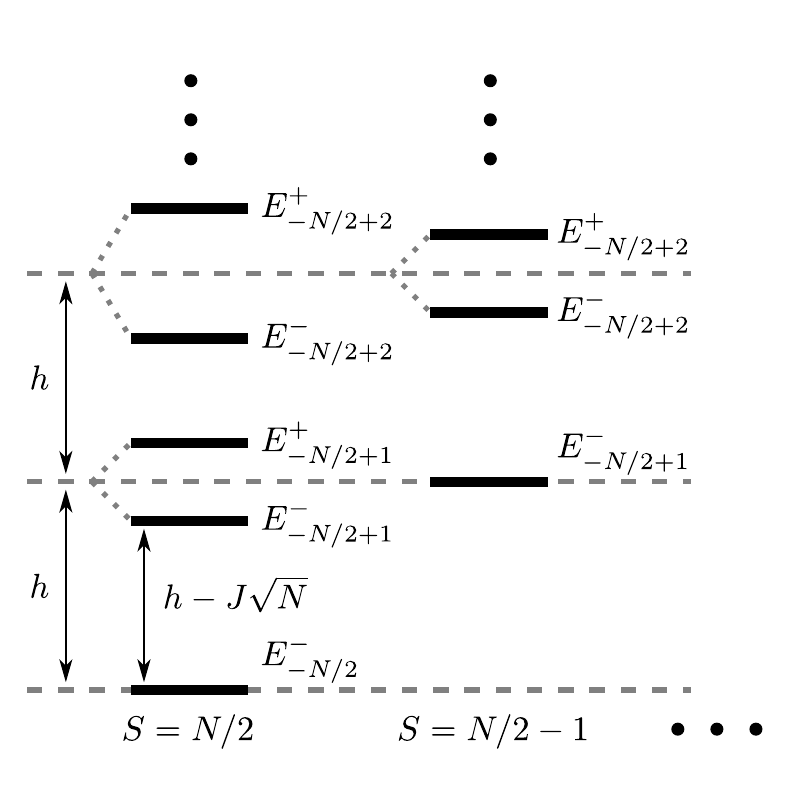}
    \caption{Low-lying eigenvalues of the spin-star Hamiltonian consisting of $N$ satellite spins for subspaces with $S=N/2$ (left column) and $S=N/2-1$ (right column).
    The thick solid lines represent the eigenvalues in ascending order from bottom to top.
    The ground and first excited state belong to the subspace with maximum angular momentum $S=N/2$.
    }
    \label{fig:spin_star_model_energy}
\end{figure}

We analytically calculate the eigensystem of the spin-star Hamiltonian with the uniform longitudinal field $h/2$.
The eigenstate of the Hamiltonian~\eqref{eq:target H with total spin} can be represented by using the quantum states $\ket{S, S^{z}}$ that are simultaneous eigenstates of the square of the total angular momentum $\hat{\boldsymbol{S}}^{2} \defas (\hat{S}^{x})^{2} + (\hat{S}^{y})^{2} + (\hat{S}^{z})^{2}$ and the $z$ component of the total angular momentum $\hat{S}^{z}$.
The variables in the quantum states take discrete values, $S = N/2, N/2 - 1, \dotsc $ and $S^{z} = -S, -S + 1, \dotsc , S - 1, S$.
Firstly, we can confirm that the eigenstates of the problem Hamiltonian for $-S + 1 \le S^{z} \le S - 1$ are given as
\begin{align}
    \ket{\psi_{S^{z}}^{+}} &= \frac{1}{\sqrt{2}} \ket{\downarrow}\ket{S, S^{z}}
                            + \frac{1}{\sqrt{2}} \ket{\uparrow}\ket{S, S^{z}-1}, \\
    \ket{\psi_{S^{z}}^{-}} &= \frac{1}{\sqrt{2}} \ket{\downarrow}\ket{S, S^{z}}
                            - \frac{1}{\sqrt{2}} \ket{\uparrow}\ket{S, S^{z}-1},
    \label{eq:spin-star eigenstate minus}
\end{align}
and the eigenvalues are
\begin{align}
    E_{S^{z}}^{\pm} = h\left( S^{z} - \frac{1}{2} \right) \pm J \sqrt{S(S+1) - S^{z}(S^{z}-1)}.
    \label{eq:HP_energies}
\end{align}
Secondly, for $S^{z} = -S$, the eigenstate is
\begin{align}
    \ket{\psi_{-S}^{-}} = \ket{\downarrow}\ket{S, -S}
\end{align}
and the eigenvalue is
\begin{align}
    E_{-S}^{-} = -h \left(S + \frac{1}{2}\right)
    \label{eq:HP_ground_energy}
\end{align}
Finally, the eigenstate for $S^{z} = S$ is given as
\begin{align}
    \ket{\psi_{S}^{+}} = \ket{\uparrow}\ket{S, S},
\end{align}
and the eigenvalue is
\begin{align}
    E_{S}^{+} = h \left(S + \frac{1}{2}\right).
\end{align}
Some low-lying eigenvalues of the spin-star Hamiltonian with $N$ satellite spins are shown in Fig.~\ref{fig:spin_star_model_energy}.

\section{\label{sec:setup and methods} Methods}

We compared the performance of adiabatic and non-adiabatic scheme based on the fidelity,
that is, the overlap between the target quantum state and the final state after performing the proposed schemes.
Let us consider the problem of finding the first excited state of the spin-star model.
The fidelity is defined as
\begin{align}
    F(T) = \lvert  \braket{\psi_{-S+1}^{-}|\Psi(T)} \rvert^{2},
    \label{eq:fidelity}
\end{align}
where $\ket{\psi_{-S+1}^{-}}$ is the first excited state of the spin-star model in the maximum angular momentum space $S = N/2$ given by Eq.~\eqref{eq:spin-star eigenstate minus},
and $\ket{\Psi(T)}$ is an result of QA whose evolution time is $T$.

We calculate the final state after performing the proposed schemes
by solving the Lindblad master equation.
The detail of the form of the Lindblad operator strongly depends on the choice
of the physical systems, and there are many research about the decoherence during the QA.
Here, we adopt a simple phenomenological depolarizing model described in Ref.~\cite{Hein2005entanglement}.
The Lindblad master equation used in the present paper is given as
\begin{align}
    \frac{d \rho(t)}{dt} = -i [\hat{H}(t), \rho(t)] 
    + \frac{1}{T_{1}} \sum_{i=1}^{N}\sum_{\alpha = x,y,z}\left[
        \hat{\sigma}_{i}^{\alpha}\rho(t)\hat{\sigma}_{i}^{\alpha} -\rho(t)
    \right],
    \label{eq:lindblad}
\end{align}
where $\rho(t)$ is a density matrix of the quantum state at time $t$,
$T_{1}$ is a relaxation time that corresponds to the inverse of the strength of the decoherence,
and $\hat{H}(t)$ is a total Hamiltonian of QA given as
\begin{align}
    \hat{H}(t) = \frac{t}{T} \hat{H}_{\text{P}} + \left( 1 - \frac{t}{T}\right)\hat{V}.
    \label{eq:total H}
\end{align}
Here, the problem Hamiltonian $\hat{H}_{\text{P}}$ is the Hamiltonian of the spin-star model in Eq.~\eqref{eq:target H},
and the operator $\hat{V}$ is a driver Hamiltonian that introduces quantum effects into the system,
and induces transitions between the quantum states.

In the case of the non-adiabatic scheme, we adopt the usual transverse-field term as the driver Hamiltonian, $\hat{V}/h = -\sum_{i=1}^{N}\hat{\sigma}_{i}^{x}$, where $h$ is the parameter of the longitudinal field
in Eq.~\eqref{eq:target H}.
Firstly, we prepare the ground state of the transverse-field term at the beginning of the QA,
$\ket{\Psi(0)} = \ket{{+}\dotsm{+}}$, where the quantum state $\ket{+}$ represents the eigenstate
of $\hat{\sigma}^{x}$ with the eigenvalue $+1$.
Secondly, let this state evolve by the Lindblad master equation in Eq.~\eqref{eq:lindblad}.
Finally, we calculate the fidelity given by Eq.~\eqref{eq:fidelity}.
In order to find the best evolution time that maximizes the fidelity,
we repeat the above procedure for various evolution times.
We confirmed that there exists the optimal total evolution time $T$ to maximize the fidelity.
Also, we confirmed that the fidelity converges to a finite small value in both limits of small and large $T$.

For the adiabatic scheme, we cannot use the conventional driver Hamiltonian (that is designed to find the ground state), but need to choose a suitable Hamiltonian, depending on which excited state we want to find.
More specifically, the first excited state of the conventional transverse-field term is degenerate,
which makes it difficult to achieve the target first excited state of the problem Hamiltonian
after the adiabatic scheme.
The proper superposition of the degenerate first excited state could be derived
from the backward unitary evolution of the target state based on the Hamiltonian~\eqref{eq:total H}.
However, it is generally impossible to obtain the proper superposition without knowing the target state.
In order to resolve the issue, we adopt the following inhomogeneous transverse field as the driver Hamiltonian:
\begin{align}
    \hat{V}/h = -b\hat{\sigma}_{0}^{x} - \sum_{i=1}^{N}\hat{\sigma}_{i}^{x}.
    \label{eq:driver_H_for_adiabatic}
\end{align}
Here, $h$ is the parameter of the longitudinal field in Eq.~\eqref{eq:target H}.
The parameter $b$, which takes a real value between zero and unity,
determines the energy difference between the ground and the first excited state.
This also determines the difference between the first and second excited states.
The eigenvalue of the first excited state is $E_1 = E_0 + b(E_2 - E_0)$,
where $E_0/h = -N -b$ and $E_2/h = -N -b + 2$ are the eigenvalues of the ground state and second excited state, respectively.
We can use the non-degenerate first excited state as the initial state.
Also, we let the state evolve by the Lindblad master equation in the Eq.~\eqref{eq:lindblad}
for a time $T$, and calculate the fidelity $F(T)$.
By sweeping $T$, we find the maximum fidelity with an optimized evolution time $\tau$.

In order to compare the two schemes (adiabatic and non-adiabatic one),
we consider two sets of parameters where the energy difference
between the ground and the first excited state of the target Hamiltonian varies.
The energy difference between the ground and the first excited state is $h - J \sqrt{N}$
from Eqs.~\eqref{eq:HP_energies} and \eqref{eq:HP_ground_energy}.
We adopt the strength of longitudinal field $h$ as the reference scale of energy,
and accordingly we set $h = 1$.
We consider two cases with $J/h=1/(2\sqrt{N})$ and $J/h=4/(5 \sqrt{N})$ where the ground and the first excited state are well detuned from the other excited states.
In the former case, the energy of the first excited state is much more detuned from that of the ground state than the latter case.
In the latter case, the first excited state is close to the resonance with the ground state.
We refer to the former and latter cases as \textit{detuned} and \textit{nearly-degenerate} cases,
respectively.

We used QuTiP~\cite{johansson2013qutip} to solve the Lindblad master equation~\eqref{eq:lindblad},
multidimensional arrays of NumPy~\cite{oliphant2006guide} as containers of data,
and the Brent's method implemented in SciPy~\cite{scipy}
to find the maximum fidelity.
All the figures for numerical results were generated using matplotlib~\cite{hunter2007}.

\section{\label{sec:results}Results and discussion}

\begin{figure}[htbp]
    \begin{minipage}[b]{.495\linewidth}
        \subfloat[Detuned case\label{fig:F_vs_T1_far}]{
            \centering
            \includegraphics[width=3.2truein]{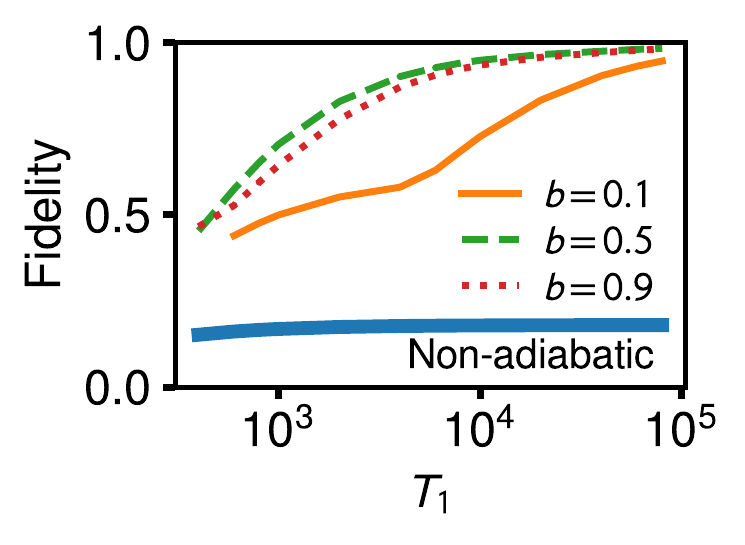}
        }
    \end{minipage}
    \begin{minipage}[b]{.495\linewidth}
        \subfloat[Nearly-degenerate case\label{fig:F_vs_T1_close}]{
            \centering
            \includegraphics[width=3.2truein]{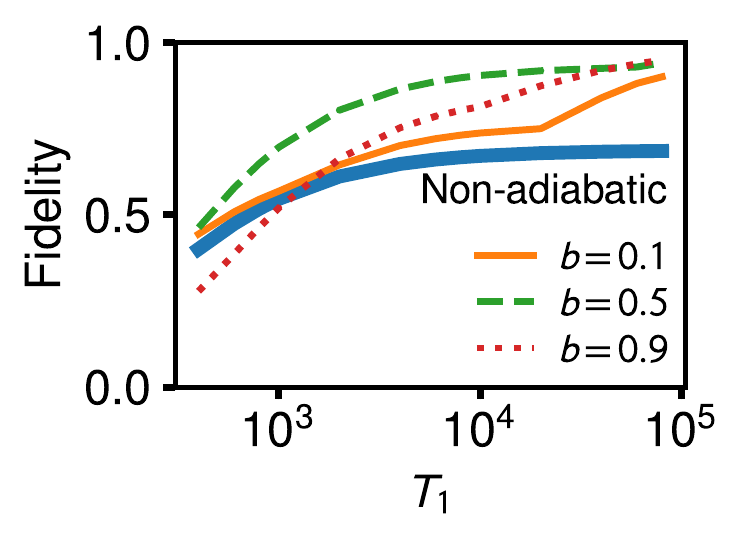}
        }
    \end{minipage}
    \caption{Maximum fidelity as a function of relaxation time $T_1$
    for (a) detuned and (b) nearly-degenerate case.
    The parameter of the longitudinal field $h$ is chosen to be unity, $h=1$,
    and $T_1$ is normalized by $1/h$.
    The normalized strength of the flip-flop interaction $J/h$ is set to $1/(2\sqrt{N})$
    for the detuned case, and $4/(5\sqrt{N})$ for the nearly-degenerate case.
    The thick solid line represents the maximum fidelity obtained from the non-adiabatic scheme.
    The thin lines show the maximum fidelities for the adiabatic scheme with the parameter
    $b=0.1$ (solid), $0.5$ (dashed), and $0.9$ (dotted).
    }
    \label{fig:F_vs_T1}
\end{figure}

Firstly, we show the maximum fidelity as a function of the relaxation time $T_{1}$, which is normalized by $1/h$.
The total number of the spins is six, consisting of five satellite spins and one central spin.
Figure~\ref{fig:F_vs_T1_far} represents the results for the detuned case.
The maximum fidelity increases as the relaxation time increases for all the data.
Regardless of the value of $b$, the maximum fidelities for the adiabatic scheme
are greater than the fidelity for the non-adiabatic scheme.
While the fidelity for the non-adiabatic scheme does not reach unity,
those for the adiabatic scheme converge to unity as the relaxation time gets longer,
so that we can obtain the target state with a high probability by using the adiabatic scheme.
In contrast, Fig.~\ref{fig:F_vs_T1_close} shows that the maximum fidelities for the adiabatic scheme are higher than those for the non-adiabatic scheme only in the regime of long relaxation time $T_{1}$ in the nearly-degenerate case.

The adiabatic scheme can achieve a higher fidelity than the non-adiabatic scheme in the regime of long relaxation time for both detuned and nearly-degenerate cases.
This is because the effect of noise is small, resulting in a time evolution that is close to the adiabatic evolution.

In the short relaxation time regime, the performance depends on the detuning of the problem Hamiltonian.
The maximum fidelity of the adiabatic scheme is always higher
than that of the non-adiabatic scheme for the detuned case (see Fig.~\ref{fig:F_vs_T1_far}).
In contrast, for the nearly-degenerate case, the fidelities for the adiabatic and non-adiabatic schemes are comparable in the short relaxation time regime.
This behavior can be interpreted from the size of the energy gap
between the instantaneous first excited state and the ground state.
In the detuned case, the energy gap is greater than the nearly-degenerate case.
The large energy gap suppresses the non-adiabatic transition from the ground to the first excited state, and causing the low fidelity for the non-adiabatic scheme.
In the nearly-degenerate case, it is possible to move the large amount of population
in the ground state to the first excited state by adjusting the evolution time $T$.
Although the energy gap for the nearly-degenerate case is smaller than the detuned case,
the adiabatic scheme can obtain as high fidelity as the non-adiabatic scheme.

\begin{figure}[htbp]
    \begin{minipage}[b]{.495\linewidth}
        \subfloat[Detuned case\label{fig:tau_vs_T1_far}]{
            \centering
            \includegraphics[width=3.2truein]{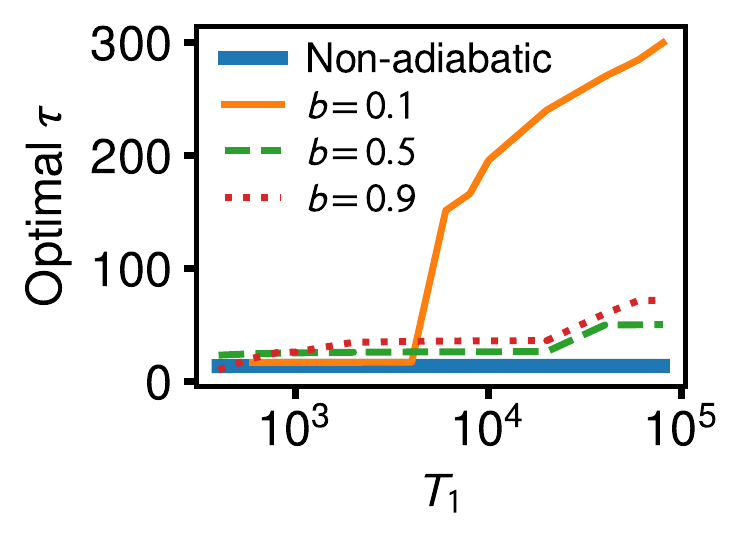}
        }
    \end{minipage}
    \begin{minipage}[b]{.495\linewidth}
        \subfloat[Nearly-degenerate case\label{fig:tau_vs_T1_close}]{
            \centering
            \includegraphics[width=3.2truein]{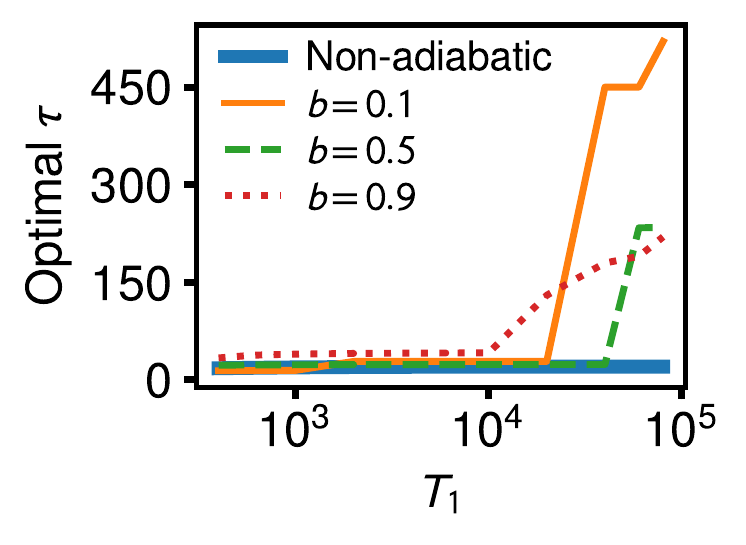}
        }
    \end{minipage}
    \caption{Optimal evolution time $\tau$ as a function of relaxation time $T_1$
    for (a) detuned and (b) nearly-degenerate case.
    Each time is normalized by $1/h$.
    We choose the same parameters $h$ and $J$ as those in Fig.~\ref{fig:F_vs_T1}.
    The thick solid line represents the optimal evolution time for the non-adiabatic scheme.
    The thin lines show the optimal evolution time for the adiabatic scheme
    with the various values of $b$,
    namely $0.1$ (solid), $0.5$ (dashed), and $0.9$ (dotted).
    }
    \label{fig:tau_vs_T1}
\end{figure}

Secondly, we plot the optimized evolution time $\tau$ where the fidelity is maximized,
where $\tau$ is normalized by $1/h$.
Figure~\ref{fig:tau_vs_T1_far} (\ref{fig:tau_vs_T1_close}) corresponds to the detuned (nearly-degenerate) case.
We can see that the optimal evolution time $\tau$ for the non-adiabatic scheme
in order to get the best fidelity is almost constant for both cases,
whereas the optimal evolution time $\tau$ for each adiabatic scheme increases as the relaxation time $T_1$ increases.

The optimal evolution time $\tau$ in units of $1/h$ for the non-adiabatic scheme
is of the order of 10, which is much smaller than the relaxation time that we investigated.
Since the induced noise hardly affects the system in this regime,
the optimal evolution time is almost independent of $T_{1}$.

The optimal evolution time for the adiabatic scheme strongly depends on $b$.
This behavior is determined by the relaxation time $T_1$ and the minimum energy gap between the ground state and the other states.
Let us focus on the detuned case.
We confirmed that the Hamiltonians for the cases with $b=0.5$ and $0.9$
have sufficiently large energy gaps, and hence the evolution time $T$ to satisfy
the adiabatic condition is much smaller than the relaxation time $T_1$ in each case.
Since the adiabatic dynamics is dominant, the fidelity increases as the evolution time increases until the decrease of the fidelity by decoherence overtakes the increase by the adiabatic dynamics.
For this reason, the optimal evolution time $\tau$ increases as the relaxation time increases.
In contrast, the adiabatic condition in the case of $b=0.1$ is satisfied only for the long relaxation time.
In the short relaxation time regime, the adiabatic scheme with $b=0.1$ achieves a high fidelity
using the interference of the quantum state; that is, the dynamics is not adiabatic anymore.
We describe the dynamics in detail in Appendix~\ref{sec:dynamics_of_population}.
In the nearly-degenerate case, the energy gap near the end of the QA process is smaller than
that in the detuned case.
Hence, longer relaxation time is needed for the adiabatic dynamics compared to the detuned case.
This is why the flat region of the optimal evolution time grows
when we simulate the adiabatic scheme for the nearly-degenerate case.
In order for the success of the adiabatic scheme, long relaxation time is required.

\section{\label{sec:conclusion}Conclusion}

The major aim of the present paper is to propose a scheme to search \textit{any} excited states in the problem Hamiltonian.
Also, we show the advantage of the adiabatic scheme
against the non-adiabatic scheme for the first excited state search.
To quantify the performance, we adopt a fidelity between the state obtained from the dynamics
and the first excited state of the target Hamiltonian.
As an example to quantify the performance, we consider the spin-star model that contains flip-flop
interaction terms, which provides us with a non-trivial excited state unlike the Ising Hamiltonian.
To take the noise into account, we adopt a Lindblad master equation
that represents the depolarizing noise.
We implemented the numerical simulation to study the performance of each scheme.
We conclude that the adiabatic scheme can achieve a higher fidelity than the non-adiabatic scheme
if the energy gap between the target energy level and the other levels is large,
while the fidelity of the adiabatic scheme is still comparable with that of the non-adiabatic scheme
even when such a gap becomes small.

Such a search of the excited state is one of the important topics of quantum chemistry.
In order to apply our proposal to the excited state search,
it is necessary to realize the non-stoquastic Hamiltonian in QA device,
since the Hamiltonian for the quantum chemistry calculations can include $XX$ interactions
as well as $ZZ$ interactions that cause the negative sigh problem in the quantum Monte Carlo method.
In the current technology, only a few qubit can be coupled each other
via the non-stoquastic Hamiltonian~\cite{ozfidan2020demonstration,ferguson2017non}.
However, it is expected that more sophisticated technique of the fabrication
and the integration would be applied to the qubits in the future
so that we could use a better device for the QA,
which will lead us to realize the practical applications of the quantum chemistry using QA.
It is also worth mentioning that D-Wave machine uses only DC magnetic fields,
and accordingly it cannot prepare the initial excited state by applying microwave $\pi$ pulses.
On the other hand, there is another type of QA that uses microwave (or radio frequency) driving
fields such as Refs.~\cite{matsuzaki2020quantum,goto2016bifurcation},
and this kind of architecture is suitable for our scheme.
Our theoretical proposals would be useful to provide one of the examples to check
whether such future QA devices outperform the classical computers.

\begin{acknowledgments}
This paper was partly based on results obtained from a project commissioned by the New Energy and Industrial Technology Development Organization (NEDO), Japan.
This work was also supported by Leading Initiative for Excellent Young Researchers MEXT Japan,
MEXT KAKENHI (Grant No.~15H05870), and JST, PRESTO (Grant No.~JPMJPR1919) Japan.
The authors would like to acknowledge Dr.\ Shu Tanaka and Dr. Hideaki Hakoshima
for fruitful discussion and useful information relevant to this study.
\end{acknowledgments}

\appendix

\section{\label{sec:dynamics_of_population}Dynamics of the population in the adiabatic scheme}

\begin{figure}[htbp]
    \begin{minipage}[b]{.495\linewidth}
        \subfloat[$T_1 = 10^3$\label{fig:F_vs_T_far_T1_1000}]{
            \centering
            \includegraphics[width=3.2truein]{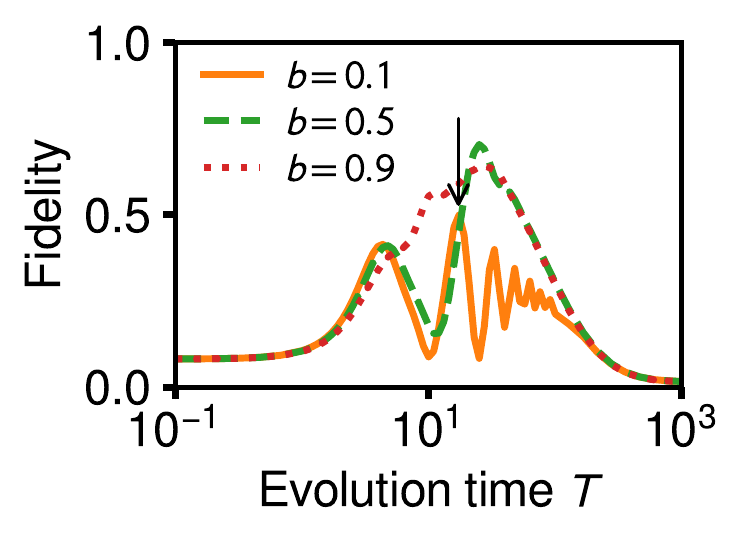}
        }
    \end{minipage}
    \begin{minipage}[b]{.495\linewidth}
        \subfloat[$T_1 = 10^4$\label{fig:F_vs_T_far_T1_10000}]{
            \centering
            \includegraphics[width=3.2truein]{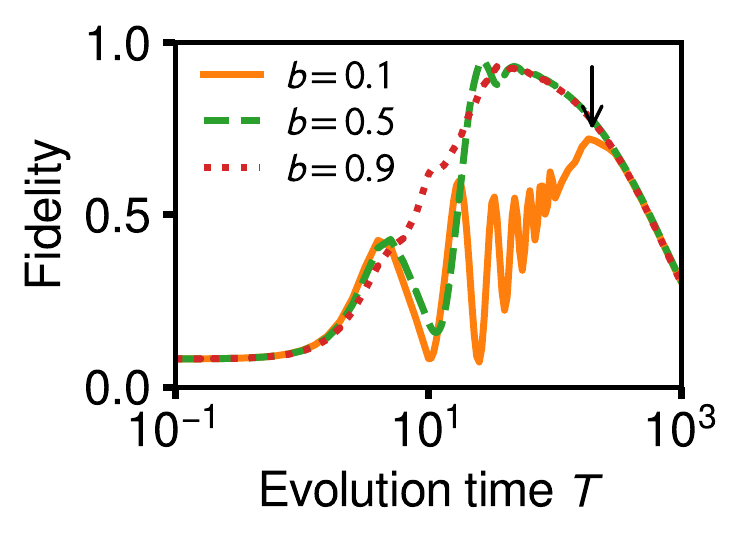}
        }
    \end{minipage}
    \caption{Fidelity as a function of the evolution time $T$ for the adiabatic scheme
    in the detuned case with the relaxation times:
    (a) $T_1 = 10^3$ and (b) $T_1 = 10^4$.
    Here, $T$ and $T_1$ is normalized by $1/h$, and we set $h=1$.
    The fidelities for each value of $b$ in the driver Hamiltonian
    in Eq.~\eqref{eq:driver_H_for_adiabatic} are represented by the solid ($b=0.1$),
    the dashed ($b=0.5$), and the dotted line ($b=0.9$).
    The maximum peak for $b=0.1$ is denoted by the down arrow in each figure.
    }
    \label{fig:F_vs_T_far}
\end{figure}

In this appendix, we explain the dynamics of the adiabatic scheme especially for the detuned case
in order to understand how to obtain a high fidelity using the scheme.
Figure~\ref{fig:F_vs_T_far} shows the fidelity for various values of the evolution time $T$
in units of $1/h$.
We can find that the fidelity decays for long evolution time due to noise.
The fidelity peak for each case $b=0.5$ and $0.9$ is located just before the decay.
The peaks of the maximum fidelities for $b= 0.5$ and $0.9$
shift to the right as the relaxation time $T_1$ normalized by $1/h$ increases from $10^3$ to $10^4$.
The fidelity is less affected for longer relaxation time, and hence the optimal evolution time $\tau$ for long relaxation time is longer than that for short relaxation time.

\begin{figure}[htbp]
    \begin{minipage}[b]{.495\linewidth}
        \subfloat[$T_1 = 10^3$\label{fig:P_vs_T_far_T1_1000}]{
            \centering
            \includegraphics[width=3.2truein]{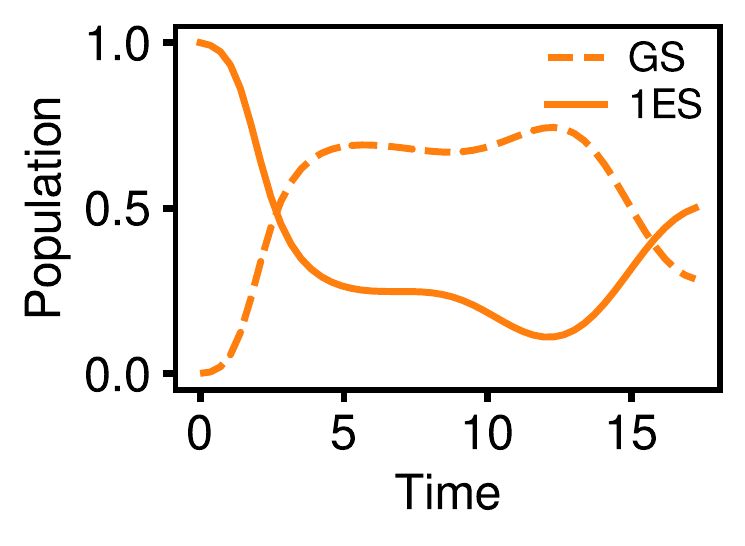}
        }
    \end{minipage}
    \begin{minipage}[b]{.495\linewidth}
        \subfloat[$T_1 = 10^4$\label{fig:P_vs_T_far_T1_10000}]{
            \centering
            \includegraphics[width=3.2truein]{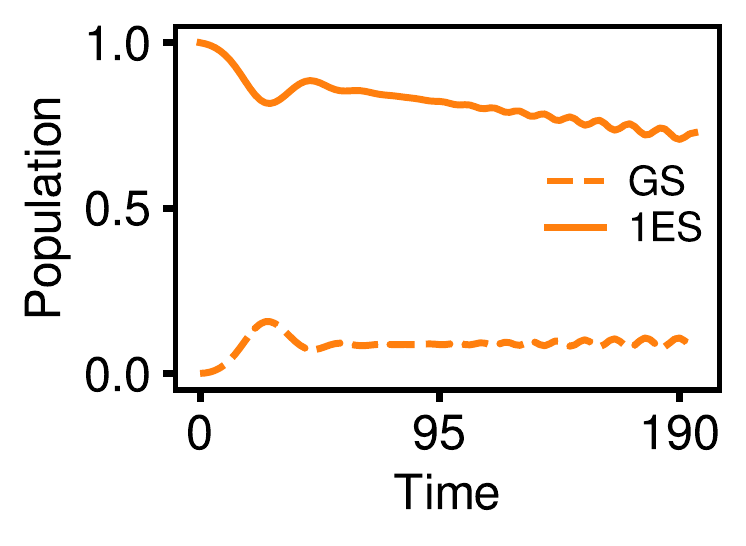}
        }
    \end{minipage}
    \caption{Populations of the instantaneous energy eigenstates during the time evolution by the adiabatic scheme in the detuned case with $b=0.1$ at the optimal evolution time $\tau$ for each relaxation time:
    (a) $T_1 = 10^3$ and (b) $T_1 = 10^4$.
    Each time is normalized by $1/h$, and we set $h=1$.
    The optimal evolution times are (a) $\tau = 17.21$, and (b) $\tau = 196.0$,
    which are denoted by the down arrows in Fig.~\ref{fig:F_vs_T_far}.
    The population of the instantaneous ground state (GS) is shown by the dashed line,
    and the instantaneous first excited state (1ES) is denoted by the solid line.
    }
    \label{fig:P_vs_T_far}
\end{figure}

In contrast, when the evolution time $T$ to satisfy the adiabatic condition is longer than the relaxation time $T_1$, the optimal evolution time $\tau$ is determined by non-trivial dynamics of the initial state.
Some populations in the first excited state transfer to the ground state,
and then return to the first excited state during the time evolution when we ingeniously choose an evolution time $T$.
The fidelity for $b=0.1$ has multiple peaks around some evolution times as clearly shown in Fig.~\ref{fig:F_vs_T_far}.
We found that, as shown in Fig.~\ref{fig:P_vs_T_far_T1_1000},
the revival of the population of the first excited state actually happened
in the regime of short relaxation time when the optimal evolution time
for $b=0.1$ is almost constant in Fig.~\ref{fig:tau_vs_T1_far}.
Although the adiabatic scheme for $b=0.1$ seems to be able to achieve the high fidelities,
it is difficult to choose the optimal evolution time $\tau$ because the fidelity rapidly decays
as the evolution time deviates from the optimal point.
Strictly speaking, for the regime where the optimal evolution time $\tau$ shown in Fig.~\ref{fig:tau_vs_T1_far} is almost constant with $b=0.1$, the dynamics is not adiabatic anymore.
When the relaxation time $T_1$ is sufficiently long, the dynamics is close to the adiabatic.
Actually, in Fig.~\ref{fig:P_vs_T_far_T1_10000}, there is no revival
between the ground and first excited state, which is typical for the adiabatic dynamics.

\bibliography{reference}

\end{document}